# UAV-Based in-band Integrated Access and Backhaul for 5G Communications

Abdurrahman Fouda[†], Ahmed S. Ibrahim[†], Ismail Guvenc[‡] and Monisha Ghosh[*]
[†] Department of Electrical and Computer Engineering, Florida International University, Miami, FL, Email: {afoud004, aibrahim}@fiu.edu
[‡] Department of Electrical and Computer Engineering, North Carolina State University, Raleigh, NC, Email: iguvenc@ncsu.edu
[*] Department of Computer Science, University of Chicago, Chicago, IL, Email: monisha@uchicago.edu

*Abstract*—We introduce the concept of using unmanned aerial vehicles (UAVs) as drone base stations for in-band Integrated Access and Backhaul (IB-IAB) scenarios for 5G networks. We first present a system model for forward link transmissions in an IB-IAB multi-tier drone cellular network. We then investigate the key challenges of this scenario and propose a framework that utilizes the flying capabilities of the UAVs as the main degree of freedom to find the optimal precoder design for the backhaul links, user-base station association, UAV 3D hovering locations, and power allocations. We discuss how the proposed algorithm can be utilized to optimize the network performance in both large and small scales. Finally, we use an exhaustive search-based solution to demonstrate the performance gains that can be achieved from the presented algorithm in terms of the received signal to interference plus noise ratio (SINR) and overall network sum-rate.

## I. Introduction

The use of low-altitude unmanned aerial vehicles (UAVs) as flying network entities in cellular communications has recently attracted increasing attention from industry [1] and academia [2]–[4]. By integrating UAVs as drone base stations (BSs), significant improvements can be made in the coverage and connectivity [5]–[7] or the capacity [8], [9] of wireless networks. Flying platforms are considered as a potential cost and energy efficient solution for 5G networks, where on one hand they offer the flexibility to be integrated in fast cellular deployments and on the other hand they can provide high data rate coverage areas with low transmit power due to their effective line-of-sight (LOS) capabilities.

In contrast to conventional macro-cell deployments, where optical fiber is considered as an appropriate medium for the transport network traffic, transport transmissions in the two-tier drone cellular networks depend on either wireless backhaul (BH) connections if the UAVs are utilized as drone BSs, or on the fronthaul connection, if the UAVs are utilized as drone remote radio heads (RRH). UAVs can be utilized as drone relays similar to the functional splitting in cloud RAN (C-RAN) scenarios between the central units (CUs) and distributed units (DUs) [10]. Integrated Access and Backhaul (IAB) has recently emerged as a potential solution for flexible and massive deployments of the 5G New Radio (NR) without densifying the terrestrial infrastructure proportionately [11].

This work is supported in part by the National Science Foundation under Awards No. CNS-1618692 and CNS-1618836.

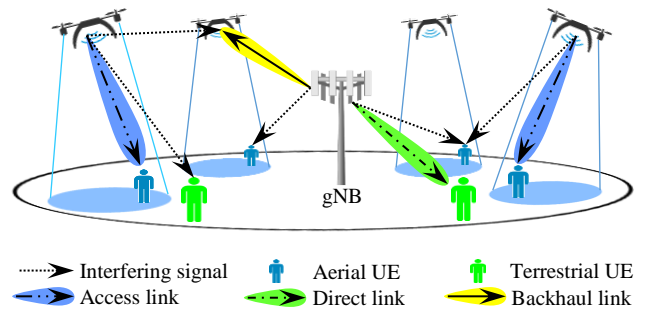

Fig. 1. In-band integrated access and backhaul (IB-IAB) system architecture for multi-tier 5G networks: UAVs can be users themselves, or operate as drone BSs to serve other users.

The concept of IAB allows the next generation NodeB (gNB) to share same spectral resources between access and BH links in order to meet the significant increase in the wireless traffic and data-demand in 5G networks [12], [13]. The problem of user association with drone BSs has been studied in [2], [4], [14] without considering the BH constraints. To the best of our knowledge, this paper is the first in literature to consider the integration of UAVs as drone BSs in an IAB scenario.

In this paper, we propose a system that utilizes UAVs as drone BSs in an IAB scenario for 5G networks. We propose an optimization framework to find the optimal precoder design for BH links, user-base station association, UAV 3D hovering locations and power allocation for forward link transmissions. We utilize the proposed framework to optimize the network performance from both large and small-scale perspectives. The rest of this paper is organized as follows. In Section II we present a system model for forward link transmissions in an in-band IAB (IB-IAB) multi-tier drone scenario. The problem formulation is presented in Section III, followed by results in Section IV that demonstrate the performance gains that can be acquired from the proposed system model. Finally, conclusions are drawn in Section V.

## II. System Model

We consider a forward link transmission scenario in an IB-IAB multi-tier drone cellular network as shown in Fig. 1. The first tier includes a gNB that on one hand provides access links to the terrestrial UEs (tUE) and on the other hand provides BH links to UAVs. The second tier represents the UAVs and the aerial UEs (aUE) associated with them. The operation mode is in-band frequency division duplex that utilizes the available resources efficiently. Fig. 1 also clarifies the definitions that are utilized throughout the paper to refer to the proposed system model. We use *forward link* to denote data transmission from *gNB to tUE, UAV to aUE* and *gNB to UAV*. An *IB-UAV* uses the same spectral resources for BH and access links, while an *IAB-gNB* uses same spectral resources for direct and BH links.

We exploit massive MIMO capabilities at gNB and utilize Multi-User MIMO (MU-MIMO) to fully reuse the available bandwidth between the direct and BH links. Frequency division multiplexing is used for forward link transmissions of access and direct links. We invoke the 5G heterogenous network (HetNet) archititure in [12] and integrate the UAVs as drone BSs in an IAB multi-tier drone HetNet.

Let $\mathcal{D} = \{1, \ldots, D\}$ represent the set of UAVs where $|\mathcal{D}| = D$. Each UAV is equipped with single receiving antenna and $N_d^{Tx}$ transmitting antennas. The gNB $g$ is equipped with $N_g^{Tx}$ transmitting antennas. The set $\mathcal{U} = \{1, \ldots, U\}$ represents the total number of UEs in the simulated scenario where $|\mathcal{U}| = U$. Both tUEs and aUEs are equipped with a single receiving antenna. $\mathcal{M} = \{1, \ldots, M\}$ is the group of $D$ BH links and the direct link of the gNB forward link transmissions to the tUEs where $M = D + 1$. $\mathcal{M}$ also denotes the total number of BSs in the simulated scenario. $\mathcal{T} = \{1, \ldots, T\}$ is the set of associated tUEs and UAVs to the gNB, where $|\mathcal{T}| = T$. The association vector that defines the serving UAV for each aUE is interpreted as $\boldsymbol{a} \in \mathbb{R}^{1 \times D}$ where $\boldsymbol{a} = [a_1 \ldots a_D]$ with $a_d$ is the index of the UAV that an aUE is associated with and $d \in \mathcal{D}$.

*A. Channel Model*

*1) Non-terrestrial links*

We invoke the pathloss models in [2], [15] to model the air-to-ground communication links. The path loss reciprocity between access (ATG) and BH (GTA) links is assumed to hold as both links are operating at the same spectral resources. The received power at aUE $u$ from UAV $a_d$ is $P_{a_d,u}^u(\boldsymbol{c}_d)$ which is a function of the association vector and UAV location and can be expressed as $P_{a_d,u}^u = P_{a_d,u}/\tilde{L}_{a_d,u}$, where $P_{a_d,u}$ is the assigned power for forward link transmissions and $\tilde{h}_{a_d,u} = 1/\tilde{L}_{a_d,u}$ is the average channel gain. We adopt the Rician channel model from [16], [17] to model the forward link transmissions of the BH and access links. We denote the wideband ATG multiple-input-single-output (MISO) channel between UAV $a_d$ and aUE $u$ and the GTA MISO channel between the gNB and UAV $a_d$ as $\boldsymbol{h}_{a_d,u} \in \mathbb{C}^{1 \times N_d^{Tx}}$ and $\boldsymbol{h}_{g,a_d} \in \mathbb{C}^{1 \times N_g^{Tx}}$ respectively.

*2) Terrestrial links*

We adopt the pathloss model for direct links between the gNB and tUEs from [18]. The received power at tUE $u$ can be expressed as, $P_{g,u}^u = P_{g,u}/\tilde{L}_{g,u}$, where $\tilde{h}_{g,u} = 1/\tilde{L}_{g,u}$. The wideband Rayleigh MISO channel between the gNB and tUE is represented by $\boldsymbol{h}_{g,u} \in \mathbb{C}^{1 \times N_b^{Tx}}$. We exploit the MU-MIMO capabilities at the gNB, where the linear zero-forcing beam forming (LZFBF) is designed to mitigate the *intra-tier* interference between the direct links and the BH links. The LZFBF precoder and full rank matrix at the gNB are expressed as $\boldsymbol{V}_g = \boldsymbol{H}_g^\dagger = \boldsymbol{H}_g^*[\boldsymbol{H}_g\boldsymbol{H}_g^*]^{-1} \in \mathbb{C}^{N_g^{Tx} \times M}$, where $\boldsymbol{H}_g = [\boldsymbol{h}_{g,1}\, \boldsymbol{h}_{g,2} \ldots \boldsymbol{h}_{g,M}]^T \in \mathbb{C}^{M \times N_g^{Tx}}$ is generated using the channel state information (CSI) of the MU-MIMO channels between $N_b^{Tx}$ antennas of gNB and $M$ reception points.

*B. IB-IAB Forward Link Transmissions*

The received signal at an aUE, $u$, from a UAV, $a_d$, can be expressed as the summation of the received signal, intra-tier and inter-tier interference as:

$$y_{a_d,u} = \sqrt{P_{a_d,u}^u} h_{a_d,u} x_{a_d,u} + \sum_{j \in \boldsymbol{a} \setminus a_d} \sum_{i \in \overline{\mathcal{U}}_j^a} \sqrt{P_{j,i}^u} h_{j,u} x_{j,i} + \sum_{k \in \boldsymbol{a} \cup \overline{\mathcal{U}}_g^t} \sqrt{P_{g,k}^u} \boldsymbol{h}_{g,u} \boldsymbol{v}_{g,k} x_{g,k} + n_u. \quad (1)$$

In (1), $x_{a_d,u}$ represents the data symbol transmitted from UAV $a_d$ to aUE $u$. $\overline{\mathcal{U}}_j^a$ and $\overline{\mathcal{U}}_g^t$ are the sets of aUEs associated with UAV $j$ and tUEs that are scheduled on the same frequency resources as aUE $u$ and create interference. The last term represents the inter-tier interference from BH and direct links. $\boldsymbol{v}_{g,k}$ and $P_{g,k}^u$ denote the precoding vector and the received power at aUE $u$ due to the transmission from gNB to reception point $k$. $n_u \sim \mathcal{CN}(0, \sigma)$ is the zero-mean complex Gaussian noise with power $\sigma^2$ at aUE $u$. The received signal at UAV $a_d$ from gNB can be expressed as the summation of the received signal, self-interference, inter-tier and intra-tier interference as:

$$y_{g,a_d} = \sqrt{P_{g,a_d}^{a_d}} \boldsymbol{h}_{g,a_d} \boldsymbol{v}_{g,a_d} x_{g,a_d} + \sum_{i \in \mathcal{U}_{a_d}^a} \sqrt{P_{a_d,i}^{Rx,a_d}} h_{a_d,a_d} x_{a_d,i} + \sum_{j \in \boldsymbol{a} \setminus a_d} \sum_{i \in \mathcal{U}_j^a} \sqrt{P_{j,i}^{a_d}} h_{j,a_d} x_{j,i} + \sum_{k \in \boldsymbol{a} \setminus a_d \cup \mathcal{U}_g^t} \sqrt{P_{g,k}^{a_d}} \boldsymbol{h}_{g,a_d} \boldsymbol{v}_{g,k} x_{g,k} + n_s. \quad (2)$$

In (2), $\mathcal{U}_{a_d}^a$ is the set of aUEs that are associated with UAV $a_d$, $P_{a_d,i}^{Rx,a_d}$ is the received power at the receiving antenna of UAV $a_d$ due to the forward link transmissions of the UAV itself to its associated UEs, and $\mathcal{U}_j^a$ is the set of aUEs that are associated with UAV $j$. The last term represents the intra-tier interference from BH and direct links. Finally, the received signal at tUE $u$ from gNB can be expressed as the summation of the received signal, intra-tier and inter-tier interference as:

$$y_{g,u} = \sqrt{P_{g,u}^u} \boldsymbol{h}_{g,u} \boldsymbol{v}_{g,u} x_{g,u} + \sum_{j \in \boldsymbol{a}} \boldsymbol{h}_{g,u} \boldsymbol{v}_{g,j} x_{g,j} + \sum_{j \in \boldsymbol{a}} \sum_{i \in \overline{\mathcal{U}}_j^a} \sqrt{P_{j,i}^u} h_{j,u} x_{j,i} + n_u. \quad (3)$$

The instantaneous received SINR at each reception point can be written as (4), (5), and (6). The UAVs are assumed to be full-duplex (FD) capable drone BS which can be integrated in the IB-IAB deployments and have the capability to completely mitigate the FD self-interference. Thus, the first term can be omitted from the denominator of (5). We assume perfect CSI knowledge at the gNB, where LZFBF is utilized to completely mitigate intra-tier interference between BH links and direct link [19].

$$\gamma_{a_d,u} = \frac{P_{a_d,u}^u |h_{a_d,u}|^2}{\sum_{j \in \boldsymbol{a} \setminus a_d} |h_{j,u}|^2 \sum_{i \in \overline{\mathcal{U}}_j^a} P_{j,i}^u + \|\boldsymbol{h}_{g,u}\|^2 \left(\sum_{k \in \boldsymbol{a} \cup \overline{\mathcal{U}}_g^t} P_{g,k}^u\right) + \sigma^2}, \quad (4)$$

$$\gamma_{g,a_d} = \quad (5)$$

$$\frac{P_{g,a_d}^{a_d} |\boldsymbol{h}_{g,a_d} \boldsymbol{v}_{g,a_d}|^2}{\sum_{i \in \mathcal{U}_{a_d}^a} P_{a_d,i}^{Rx,a_d} + \sum_{j \in \boldsymbol{a} \setminus a_d} |h_{j,a_d}|^2 \sum_{i \in \mathcal{U}_j^a} P_{j,i}^{a_d} + \|\boldsymbol{h}_{g,a_d}\|^2 \left(\sum_{k \in \boldsymbol{a} \setminus a_d \cup \mathcal{U}_g^t} P_{g,k}^{a_d}\right) + \sigma^2},$$

$$\gamma_{g,u} = \frac{P_{g,u}^u |\boldsymbol{h}_{g,u} \boldsymbol{v}_{g,u}|^2}{\|\boldsymbol{h}_{g,u}\|^2 \sum_{j \in \boldsymbol{a}} P_{g,j}^u + \sum_{j \in \boldsymbol{a}} |h_{j,u}|^2 \sum_{i \in \overline{\mathcal{U}}_j^a} P_{j,i}^u + \sigma^2}. \quad (6)$$

## III. PROBLEM FORMULATION

We deploy the UAVs in a highly congested area, where the congestion causes some UEs to suffer low levels of quality of service (QoS) while other UEs do not have service due to the lack of spectral resources or poor coverage. The proposed algorithm exploits the mobile capabilities of the UAVs as the main degree of freedom (DOF) to maximize the overall system sum-rate of both aUEs and tUEs while keeping the interference levels low. The overall instantaneous sum-rate is the sum of the instantaneous rates of aUEs and tUEs at each CSI instant and can be represented as follows:

$$\Re = \sum_{j \in a} \sum_{i \in \mathcal{U}_j^a} \log_2(1 + \gamma_{j,i}) + \sum_{i \in \mathcal{U}_{g_t}^t} \log_2(1 + \gamma_{g,i}). \quad (7)$$

The master optimization problem can be formulated as follows to find the optimal 3D hovering locations of the UAVs, UE-power allocation, precoder design at BH links, and the UE-association set per each UAV and gNB:

$$max_{C,\ p,\ p_g^a,\ a, V_g} \Re, \quad (8)$$

Subject to

$$\gamma_{j,i},\ \gamma_{g,i} \geq \gamma_{th}^u,\ \forall j \in a \text{ and } i \in \mathcal{U}, \quad (8\text{-a})$$

$$Tr(P_g V_g^* V_g) \leq P_g^{max}, \quad (8\text{-b})$$

where $C \in \mathbb{R}^{3 \times D}$ is the 3D location matrix of UAVs with $c_d = [x_d, y_d, z_d]^T$ and $p \in \mathbb{R}^{1 \times U}$ where $p = [P_{m,1} \ldots \ldots P_{m,U}]$ is the UE-power allocation vector with $P_{m,u}$ being the power allocated by BS $m$ for forward link transmissions of UE $u$. $m$ can be the gNB $g$ or UAV $a_d$ based on the optimal association vector. $p_g^a \in \mathbb{R}^{1 \times D}$ where $p_g^a = [P_g^{a_1} \ldots \ldots P_g^{a_D}]$ is the UAV-power allocation vector with $P_g^{a_d}$ being power allocated by gNB for forward link transmissions of BH link of UAV $a_d$. $V_g$ is the optimal precoder design at gNB. To guarantee that a target QoS is satisfied at each UE, the received SINR at tUEs and aUEs are constrained by $\gamma_{th}^u$ (8-a). $P_g^{max}$ represents maximum transmit power of the gNB and is utilized to define the power budget constraint on the precoder design. $P_g = diag(P_{g,1}, P_{g,2}, \ldots \ldots, P_{g,M})$ denotes the transmit power allocation of gNB, where $P_{g,m}$ denotes forward link transmission power assigned by gNB to link $m$.

The master optimization problem presented in (8) cannot be considered as a single optimization problem due to the severe variations between the update time instants of each optimization variable. On one hand, the optimal $C$ and $a$ should be updated every *update instant* the network reaches a predefined user-drop rate or the QoS of certain group of UEs decreases below a predetermined level. On the other hand, the optimal $p$, $p_g^a$ and $V_g$ that yield the maximum instantaneous rate should be updated each *CSI instant*. To this end, we decompose the master optimization problem in (8) into two sub-optimization problems due to the mutual dependence between the optimization variables and their update time instants. The first sub-problem (*P-I*) defines the optimal $C$, $a$, $p$ and $p_g^a$ that yield the maximum *average* system sum-rate. The second sub-problem (*P-II*) defines the optimal $V_g$ and $p_g$ to update the gNB optimal power allocations that yield the maximum *instantaneous* sum-rate on BH and direct links. The network performance is optimized on one hand utilizing *P-I* in large-scale perspective *(i.e., every update instant)* and on the other hand utilizing *P-II* in small-scale perspective.

The sub-problem *P-I* can be formulated as the sum of the average sum-rate of aUEs and tUEs as follows:

$$\max_{C,\ p,\ p_g^a,\ a} \sum_{j \in a} \sum_{i \in \mathcal{U}_j^a} \log_2(1 + \tilde{\gamma}_{j,i}) + \sum_{i \in \mathcal{U}_g^t} \log_2(1 + \tilde{\gamma}_{g,i}),$$

Subject to $\quad (9)$

$$\tilde{\gamma}_{a_d,u} = \frac{P_{a_d,u} \tilde{h}_{a_d,u}}{\sum_{j \in a \setminus a_d} \tilde{h}_{j,u} \sum_{i \in \tilde{\mathcal{U}}_j^a} P_{j,i} + \tilde{h}_{g,u}\left(\sum_{k \in a \cup \bar{u}_g^t} P_{g,k}^u\right) + \sigma^2} \geq \gamma_{th}^u, \quad (9\text{-a})$$

$$\tilde{\gamma}_{g,u} = \frac{P_{g,u} \tilde{h}_{g,u}}{\sum_{j \in a} \tilde{h}_{j,u} \sum_{i \in \bar{\mathcal{U}}_j^a} P_{j,i} + \sigma^2} \geq \gamma_{th}^u, \quad (9\text{-b})$$

$$\tilde{\gamma}_{g,a_d} = \frac{P_{g,a_d}^{a_d} \tilde{h}_{g,a_d}}{\sum_{j \in a \setminus a_d} \tilde{h}_{j,a_d} \sum_{i \in \mathcal{U}_j^a} P_{j,i}^{a_d} + \sigma^2} \geq \gamma_{th}^{BH}\ a_d \in a, \quad (9\text{-c})$$

$$\left(\frac{P_{a_d,u}}{\bar{L}_{a_d,u}} - \frac{P_{a_{d'},u}}{\bar{L}_{a_{d'},u}}\right) > 0, \forall\ d, d' \in \mathcal{D}, d \neq d', \quad (9\text{-d})$$

$$P_{a_d,u} \in [0, P_{a_d}^{max}], P_{g,u} \in [0, P_g^{max}], \quad (9\text{-e})$$

$$c_d \in [c_d^{min}, c_d^{max}], \forall\ c \in \{x, y, z\}, \quad (9\text{-f})$$

where $P_{a_d,u}(a_d)$ is the assigned power for forward link transmissions while $\tilde{h}_{a_d,u}(c_d)$ is the average channel gain and is a function of the association vector and the UAV's location. The strict inequality in (9-d) guarantees that for a UE to be connected to a UAV, the received power from this UAV must be greater than the received power from any other UAV.

The sub-problem *P-II* can be formulated as follows:

$$max_{p_g, V_g} \sum_{j \in a} \log_2(1 + \gamma_{g,j}) + \sum_{u \in \mathcal{U}_g^t} \log_2(1 + \gamma_{g,u}), \quad (10)$$

Subject to

$$\gamma_{a_d,u} \geq \gamma_{th}^{BH},\ \gamma_{g,u} \geq \gamma_{th}^u, \quad (10\text{-a})$$

$$Tr(P_g V_g^* V_g) \leq P_g^{max}, \quad (10\text{-b})$$

where $p_g \in \mathbb{R}^{1 \times T}$, $p_g = [P_{g,1} \ldots \ldots P_{g,T}]$ is the reception point-power allocation vector with $P_{g,t}$ being the power allocated by gNB for forward link transmissions of reception point $t$, and $\gamma_{th}^{BH}$ denotes the SINR threshold at BH links.

## IV. PERFORMANCE EVALUATION

In this section we use an exhaustive search process to find the feasible set of solutions of *P-I* and present results to prove the efficiency of utilizing the UAVs as drone BSs in IB-IAB networks. Due to the limited processing capabilities of the exhaustive search process we consider a single gNB and UAV in an urban macro (UMa) scenario where a limited number of UEs are distributed over a geographical area of size 1.5 km × 1.5 km. We define two scenarios of user clustered distributions to evaluate the efficiency of utilizing the UAVs' mobile capabilities. In scenario 𝔸 users are clustered in multiple hotspots that are randomly distributed in the coverage area. In scenario 𝔹 we discuss the single hotspot case where specific number of UEs are normally distributed in a single hotspot while other UEs are randomly distributed in the remaining coverage area. The simulation parameters are summarized in Table I.

TABLE I SIMULATION PARAMETERS.

| Parameter | Value | Parameter | Value |
|---|---|---|---|
| $P_g^{max}$ | 46 dBm | $P_{a_d}^{max}$ | 36 dBm |
| $\gamma_{th}^u, \gamma_{th}^{BH}$ | 3, 10 dB | $\sigma^2$ | -104 dBm |
| System Bandwidth | 20 MHz | $f_c$ | 2 GHz |

## A. Scenario 𝔸

The 3D placement of UEs and UAVs in an example scenario is depicted in Fig. 2, where 2 UEs represent a single hotspot that is close to the gNB, while 2 other UEs represent another hotspot distant from the gNB. The proposed algorithm obtains the nearest BS-based association vector as the optimal association vectors that yeilds the maximum sum-rate. Each BS divides the power resources equally between the associated UEs with it, as each two UEs are nealry located at the same distance from their serving BS. The high probability of LOS on the UAV BH link permits the gNB to allocate minimal power resources for the forward link transmission. As shown in Fig. 3 the minimum received SINR at the BH link is above the 10 dB threshold, which concludes that high SINR levels can be received at the UAV BH links with the minimum transmission power allocations from the gNB to the BH links. This also decreases the interference levels to the forward link transmissions of the access links. Fig. 3 shows that received SINR of tUEs decreases from 29 to 19 and 17 dB respectively for the sake of increasing the SINR of the aUEs. As a result, the average received SINR is improved by 7 dB when the UAV is hovering at an altitude of 200 m, and by 6.2 dB at 500 m compared with the no-UAV case.

throughput is obtained at lower hovering altitudes. On one hand the number of served UEs is low, so the UAV does not need to increase its hovering altitude to increase its coverage area. On the other hand, the pathloss between aUEs and UAVs increases at higher hovering altitudes. From Fig. 3 and Fig. 4 we conclude that the SINR and throughput improvements are much greater than the interference obtained when UAVs are utilized as drone BSs in an IB-IAB scenario.

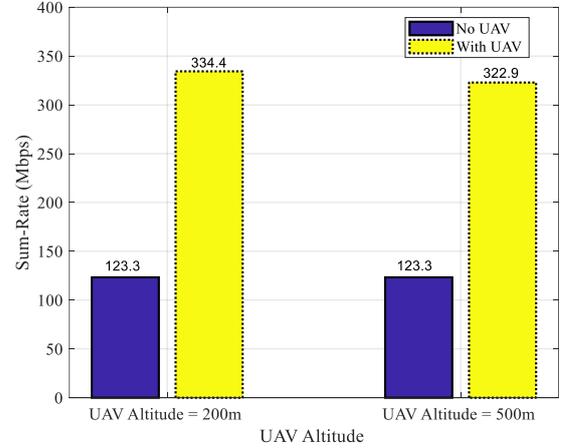

Fig. 4. Network sum-rate (Mbps) (scenario 𝔸).

## B. Scenario 𝔹

In scenario 𝔹, the UEs are clustered in a single hotspot rather than being distributed in two hotspots as shown in Fig. 5. As in the previous scenario, the optimal hovering location is obtained as the nearest location to the distant hotspot. However, we have better sum rate at 500 m when compared to the sum rate at 200 m. The UAV increases its hovering altitude to increase the coverage area and to be able to serve a larger number of UEs. As depicted in Fig. 6, the optimal received SINR at each UE is obtained at a higher hovering altitude compared with scenario 𝔸.

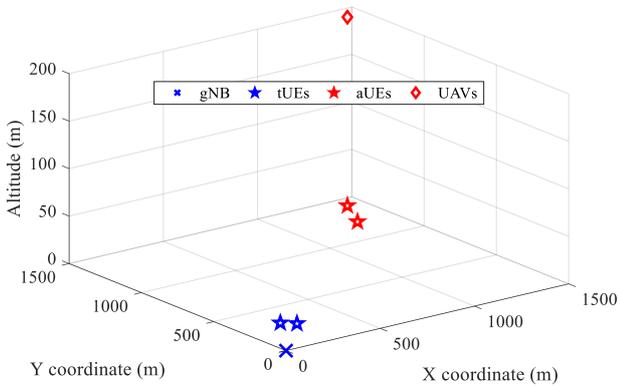

Fig. 2. UE distribution and UAV 3D placement (scenario 𝔸).

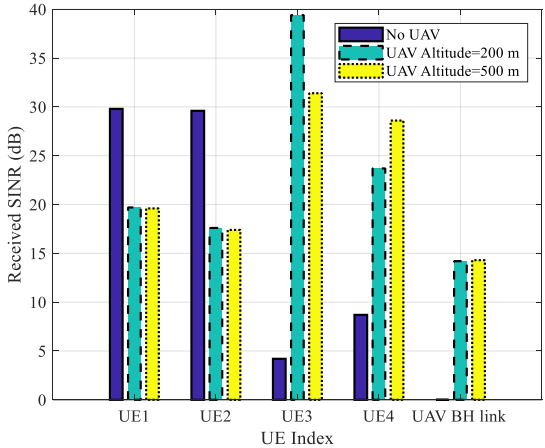

Fig. 3. Received SINR at each reception point (scenario 𝔸).

Fig. 4 illustrates that the system throughput is enhanced by 170% when the UAVs are utilized in an IB-IAB scenario compared with the no-UAV case. Higher forward link

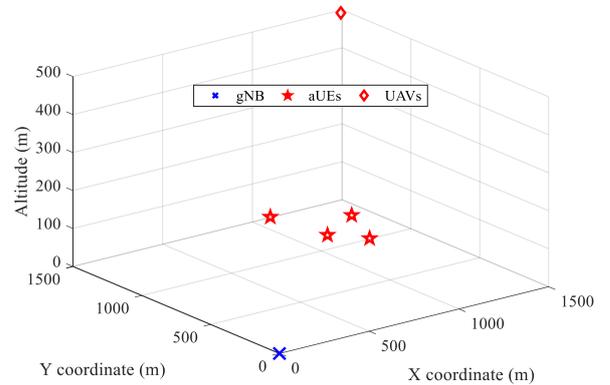

Fig. 5. UE distribution and UAV 3D placement (scenario 𝔹).

Fig. 7 illustrates that the overall network throughput is enhanced by utilizing the UAV as a drone BS. However, it is nearly half that of scenario 𝔸 as the number of connected UEs per UAV is increased and thus smaller amount of spectral resources is allocated to each UE. This concludes that the performance improvements that are obtained from utilizing the UAVs as drone BSs in IAB scenarios increase as the user heterogeneity is distributed among multiple hotspots within the coverage area, and decrease as the user heterogeneity is concentrated in a single hotspot area. Although increasing the number of UAVs that are utilized to cover this hotspot will increase the number of allocated resources to each UE in one hand, the interference levels will increase severely on the other hand decreasing received SINR per UE and overall sum-rate.

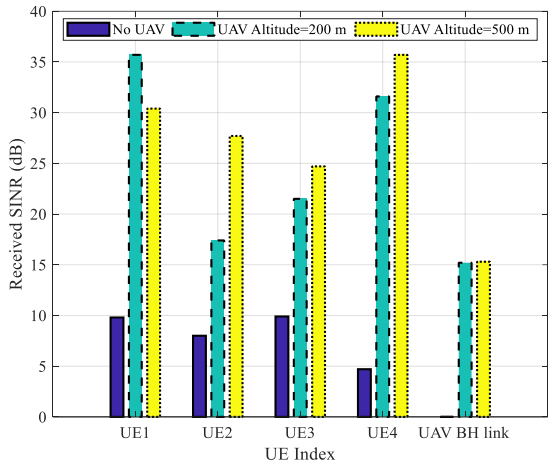

Fig. 6. Received SINR at each reception point (scenario $\mathbb{B}$).

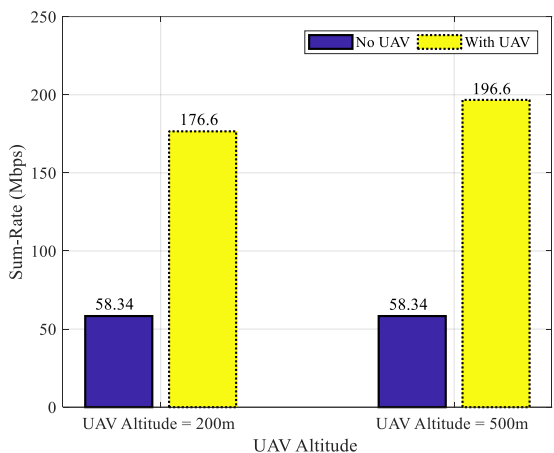

Fig. 7. Network sum-rate (Mbps) (scenario $\mathbb{B}$).

V. CONCLUSION AND FUTURE WORK

In this paper, we investigated the performance impacts of integrating UAVs as drone BSs in IB-IAB scenarios for 5G NR networks. We proposed a framework to optimize the network performance from both large and small-scale perspectives based on the congestion of the network and the received QoS per user. We demonstrated the performance gains that can be acquired from utilizing the UAVs as drone BS. The average received SINR per user is improved by 7 dB and the overall network forward link throughput is enhanced by 170%. In future work, we will derive a non-complicated optimization algorithm to optimize the network performance in both large and small scales. We will propose a learning framework to trigger the optimization algorithms of the drone-network design parameters that include the optimal user-associations, hovering locations and power allocations. sets.